\documentclass[superscriptaddress,amsmath,amssymb, aps, pra, twocolumn]{revtex4-2}

\usepackage{graphicx}
\usepackage{dcolumn}
\usepackage{bm}
\usepackage[colorlinks]{hyperref}

\begin{document}

\title{Non-inertial quantum clock frames lead to non-Hermitian dynamics}

\author{Ismael L. Paiva}
\email{ismaellpaiva@gmail.com}
\affiliation{Faculty of Engineering and the Institute of Nanotechnology and Advanced Materials, Bar-Ilan University, Ramat Gan 5290002, Israel}

\author{Amit Te'eni}
\affiliation{Faculty of Engineering and the Institute of Nanotechnology and Advanced Materials, Bar-Ilan University, Ramat Gan 5290002, Israel}

\author{Bar Y. Peled}
\affiliation{Faculty of Engineering and the Institute of Nanotechnology and Advanced Materials, Bar-Ilan University, Ramat Gan 5290002, Israel}

\author{Eliahu Cohen}
\affiliation{Faculty of Engineering and the Institute of Nanotechnology and Advanced Materials, Bar-Ilan University, Ramat Gan 5290002, Israel}

\author{Yakir Aharonov}
\affiliation{Schmid College of Science and Technology, Chapman University, Orange, California 92866, USA}
\affiliation{Institute for Quantum Studies, Chapman University, Orange, California 92866, USA}
\affiliation{School of Physics and Astronomy, Tel Aviv University, Tel Aviv 6997801, Israel}

\begin{abstract}
\textbf{Abstract.} The operational approach to time is a cornerstone of relativistic theories, as evidenced by the notion of proper time. In standard quantum mechanics, however, time is an external parameter. Recently, many attempts have been made to extend the notion of proper time to quantum mechanics within a relational framework. Here, we use similar ideas combined with the relativistic mass-energy equivalence to study an accelerating massive quantum particle with an internal clock system. We show that the ensuing evolution from the perspective of the particle's internal clock is non-Hermitian. This result does not rely on specific implementations of the clock. As a particular consequence, we prove that the effective Hamiltonian of two gravitationally interacting particles is non-Hermitian from the perspective of the clock of either particle.
\end{abstract}

\maketitle

\section*{Introduction}

Time is an intriguing physical concept that can be connected to most --- if not all --- fundamental issues in physics. A good example of that concerns how a refined understanding of time is associated with the revolution brought by the introduction of relativistic theories. In fact, Lorentz transformations, which were independently introduced by Voigt \cite{voigt1887ueber} and Lorentz \cite{lorentz1904zittingsverslag} and named as such by Poincar\'e \cite{poincare1905dynamique}, were already known for some time before Einstein's introduction of special relativity \cite{einstein1905elektrodynamik}. However, they had never been taken to their full mechanical consequences prior to Einstein's work. He did so by considering that clocks --- and rods for that matter --- are physical objects and, hence, subject to physical laws.

In quantum mechanics, the issue of time was discussed since the early days of the theory, and understandably so: while measurements are an essential element of it, it seems that the theory does not readily allow the description of measurements of time since it contains time as a parameter. One could, then, wonder about the possibility of constructing a time operator. However, arguably, much progress in this regard stagnated due to Pauli's well-known objection to such an operator \cite{pauli1933allgemeinen}, which is based on the argument that the Hamiltonian, canonically conjugate to a Hermitian time operator, would have to be unbounded from below. Although discussions about time continued to exist in the literature, to the best of our knowledge, a time operator appeared again in a discussion by Aharonov and Bohm that involved the idea of Heisenberg's cut in a special measurement scheme \cite{aharonov1961time}. Later, Garrison and Wong introduced a Hermitian time operator that measures time within a finite interval and overcomes Pauli's objection \cite{garrison1970canonically}. However, it can be argued that such a clock is nonphysical since it requires the probability of finding the system at the boundaries of the domain of the clock to vanish \cite{woods2019autonomous}. More generally, realistic physical clocks that overcome Pauli's objection can be introduced with the extension of the notion of observables from Hermitian operators to positive operator-valued measures (POVMs) \cite{holevo1982probabilistic, busch199eoperational, busch2016quantum}.

Remarkably, the canonical quantization of general relativity leads to a constraint known as Wheeler-DeWitt equation, which implies that the wave function of the universe does not evolve in time \cite{dewitt1967quantum}. For systems satisfying such a constraint, Page and Wootters introduced a formalism in which a non-interacting subsystem works as a reference for time (i.e., a clock) for the remaining parts \cite{page1983evolution}. With this scheme, which can be studied in the general context of quantum reference frames \cite{aharonov1967charge, aharonov1967observability, aharonov1984quantum, bartlett2007reference, angelo2011physics}, they showed that the usual unitary evolution given by the Schr\"odinger equation can be recovered. Their formalism has attracted much attention, especially during the last few years \cite{wootters1984time, giovannetti2015quantum, marletto2017evolution, castro2017entanglement, smith2019quantizing, giacomini2019quantum, diaz2019history, diaz2019history2, martinelli2019quantifying, castro2020quantum, smith2020quantum, ballesteros2021group, carmo2021quantifying, mendes2021non, trassinelli2022conditional, paiva2022dynamical, paiva2022flow, baumann2022noncausal, adlam2022watching}. In particular, if an arbitrary interaction term between the clock and the remaining parts is considered, a generalized Schr\"odinger equation is obtained \cite{smith2019quantizing}.

Nevertheless, the resulting evolution was found to be unitary in a vast quantity of scenarios investigated in the literature, even when gravitationally interacting clocks were considered \cite{castro2017entanglement, smith2019quantizing, castro2020quantum}. Generally, such clocks present a particular type of time dilation and can also undergo decoherence, losing their ability to behave as good clocks \cite{castro2017entanglement}. However, it is still argued that in circumstances for which they still work as clocks, the evolution of the rest of the systems from their perspective is unitary \cite{castro2020quantum}.

That said, non-unitary evolution does manifest itself in energy measurements of clock systems \cite{paiva2022flow}. More precisely, the dynamics of the rest of the system from the perspective of a clock undergoing a von Neumann measurement of energy is non-unitary even when the final ``collapse'' (or update) of the state of the system is not taken into account. Specifically, the dynamics from the perspective of the clock being measured was found to be governed by a non-Hermitian Hamiltonian. It was even speculated that such clocks are non-inertial frames of reference. However, one may question whether the clock retains its ability to measure time when its energy is being measured, putting in check the fundamental nature and the significance of the non-unitary dynamics in this case.

In this article, we show that non-Hermitian Hamiltonians are likely an unavoidable element in fully operational treatments of time in the Page and Wootters framework. More explicitly, we prove that the resulting evolution from the perspective of the proper time (i.e., internal clock) of an accelerating massive particle is generated by a non-Hermitian Hamiltonian and, generally, non-unitary. We also analyze gravitationally interacting clocks from this new perspective, explaining how to reconcile our results with previous ones, even though they may seem to be at odds. Important in our approach is a post-Newtonian correction to the mass: the mass-energy equivalence. Such a correction has previously led to other worth-noting results \cite{pikovski2015universal, castro2017entanglement, sonnleitner2018mass, zych2019gravitational, smith2019quantizing, smith2020quantum}.

\section*{Results}

\textbf{Time evolution given by quantum clocks.} Let $A$ denote a physical clock system. If $H_A$ is the system's Hamiltonian, time states are built as
\begin{equation}
    |t_A+t'_A\rangle \equiv e^{-iH_A t'_A/\hbar} |t_A\rangle.
    \label{eq:cov-rule}
\end{equation}
From these states, a time operator $T_A$ can be constructed. If different time states are orthogonal to each other, i.e., $\langle t_A|t'_A\rangle = \delta(t_A-t'_A)$ for every $t_A$ and $t'_A$, $T_A$ is Hermitian and, moreover, it is canonically conjugate to $H_A$. However, the resulting time states are not always orthogonal to each other \cite{busch199eoperational, busch2016quantum, loveridge2019relative, hohn2021trinity}. These represent more realistic clocks, with the lack of orthogonality reflecting the absence of infinite resolution of the clock. In this work, clocks are not assumed to be ideal.

Observe that, regardless of whether the time states are orthogonal to each other or not, Eq. \eqref{eq:cov-rule} implies that
\begin{equation}
    \frac{\partial}{\partial t_A} |t_A\rangle = -\frac{i}{\hbar} H_A |t_A\rangle.
    \label{eq:gen}
\end{equation}
This means that, being the generator of translations in time states, $H_A$ acts on these states as a time derivative.

Besides clock $A$, let the other relevant systems be represented by the index $M$. Also, assume that the joint system satisfies the Wheeler-DeWitt equation
\begin{equation}
    H|\Psi\rangle\rangle = 0,
    \label{eq:constraint}
\end{equation}
where $H$ is the Hamiltonian of the joint system. Here, the double-ket notation is used to identify the entire system, which does not evolve with respect to an external time. It is noteworthy that $H$ is defined as an operator acting upon $\mathcal{H}_{\text{kin}} \equiv \mathcal{H}_A\otimes\mathcal{H}_M$. However, in case of operators with continuous spectra containing zero, like in the case of $H$, the states $|\Psi\rangle\rangle$ satisfying the constraint, called physical states, are not normalized in the inner product on this space \cite{smith2019quantizing, hohn2020switch, rovelli2004quantum}. Thus, a new space $\mathcal{H}_{\text{phys}}$ is constructed for the solutions of Eq. \eqref{eq:constraint} with the inner product \cite{smith2019quantizing, hohn2020switch, rovelli2004quantum}
\begin{equation}
    \langle\langle \Psi | \Phi \rangle\rangle_{\text{phys}} \equiv \langle\langle \Psi | (|t_A\rangle\langle t_A| \otimes I_M) | \Phi \rangle\rangle.
\end{equation}

If $H_M$ denotes the Hamiltonian of the system of interest and $H_{\text{int}}$, an arbitrary interaction between $A$ and $M$, we have
\begin{equation}
    H = H_A + H_M + H_{\text{int}}.
    \label{eq:hamiltonian-generic}
\end{equation}
Moreover, define $|\psi(t_A)\rangle\equiv\langle t_A|\Psi\rangle\rangle$. As a result, $|\Psi\rangle\rangle$ can be written as
\begin{equation}
    |\Psi\rangle\rangle = \int dt_A \, |t_A\rangle \otimes |\psi(t_A)\rangle.
\end{equation}

Also, Eq. \eqref{eq:constraint} implies that $\langle t_A|H|\Psi\rangle\rangle$ vanishes, which, in turn, can be written as \cite{smith2019quantizing}
\begin{equation}
    i\hbar \frac{\partial}{\partial t_A} |\psi(t_A)\rangle = H_M |\psi(t_A)\rangle + \int dt'_A \ K(t_A,t'_A)|\psi(t'_A)\rangle,
    \label{eq:schrod-general}
\end{equation}
giving rise to a generalized Schr\"odinger equation. Here, $K(t_A,t'_A) \equiv \langle t_A|H_{\text{int}}|t'_A\rangle$. Observe that the use of Eq. \eqref{eq:constraint} in the derivation of the above expression assures that we are restricted to the subset of physical states.

In many instances, it is desirable to let $|\Psi\rangle\rangle$ be composed of various subsystems and, in particular, multiple clocks. In cases with multiple clocks, the dynamics can be studied from the perspective of any of them \cite{hohn2020switch, castro2020quantum, hohn2021trinity}.

\begin{figure*}
    \centering
    \includegraphics[width=\textwidth]{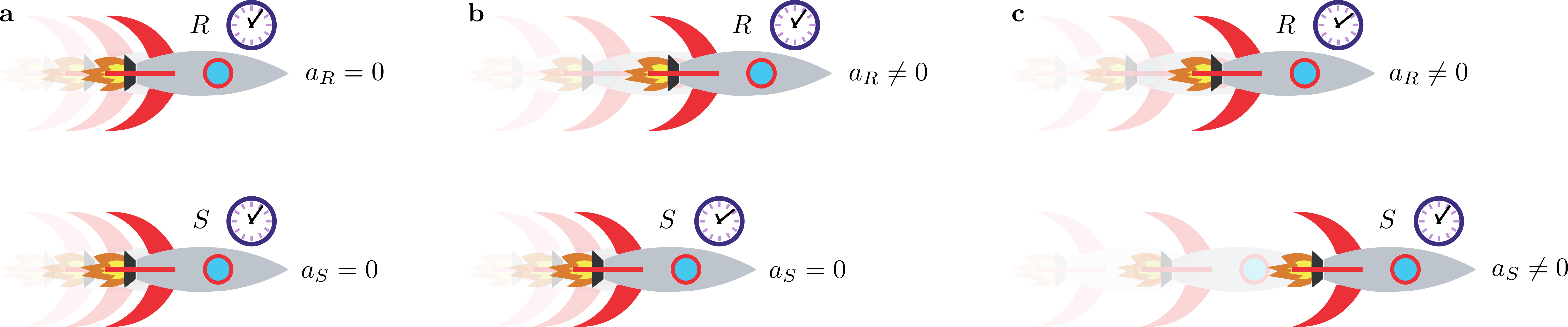}
    \caption{\textbf{Representation of non-interacting travelling rocket ships.} Ships $R$ and $S$ have each their own internal quantum time. If both rockets are at rest or moving with constant speed ($a_R=a_S=0$), as shown in \textbf{a}, the time evolution of systems described by either of them is unitary. However, if rocket $R$ starts accelerating ($a_R\neq0$), as displayed in \textbf{b}, then the time evolution of systems from its clock's perspective is, generally speaking, non-unitary, while the evolution from the perspective of rocket $S$'s clock remains unitary. Finally, none of the clocks gives a unitary evolution if both rockets are accelerating, as illustrated in \textbf{c}.}
    \label{fig:rockets}
\end{figure*}

\textbf{Accelerating clock frames.} Now, we connect the acceleration of clocks to the observance of non-Hermitian dynamics. First, a discussion of how acceleration affects the Hamiltonian of a free particle is needed. In fact, classically, this influence should be taken into account with the addition of a potential $V$. Restricting our study to potentials given by a function of the position $x$ and recalling from Newtonian physics that $ma = -dV/dx$, where $m$ is the mass of the system and $a$ is its acceleration, we conclude that
\begin{equation}
    V(x) = - m \int_{x_0}^x a(x') dx'.
\end{equation}
In the above, it was assumed for simplicity that $a$ can be parametrized by $x$ and $V(x_0)=0$. Defining $f(x) \equiv - \int_{x_0}^x a(x') dx'$, the Hamiltonian of the system becomes $H+mf(x)$.

In the quantum case, we can, then, consider a system composed by a massive particle $M$ and an internal clock $A$. Initially, we assume that they do not interact. If $H_M$ is the free Hamiltonian of $M$, $H_A$ is the Hamiltonian of the clock, and $X_M$ is the position operator associated with the center-of-mass of $M$, we can write the Hamiltonian of the system as $H = H_A + H_M + m f(X_M)$.

Now, a post-Newtonian correction can be added to $H$ by applying the mass-energy equivalence \cite{einstein1911einfluss}. For this, the mass of the system is treated as an operator and its value is updated with the Hamiltonian of the internal degrees of freedom \cite{pikovski2015universal, castro2017entanglement, sonnleitner2018mass, zych2019gravitational, smith2019quantizing, smith2020quantum}. In our case, it implies that the mass of the systems becomes $m+H_A/c^2$. Then, redefining $f$ to absorb the constant $1/c^2$, we can write
\begin{equation}
    H = H_A + H_M + H_A f(X_M).
    \label{eq:ham-accel-gen}
\end{equation}
For simplicity, we have neglected, as we also do in the rest of the article, the term with the mass $m$ and have considered only $H_A/c^2$. This is done to remove extra terms that would still be associated with a unitary dynamics. The mass $m$ can be added back without affecting the analysis presented here. Moreover, observe that the potential (associated with $f$) couples to internal degrees of freedom of the system \cite{pikovski2015universal}.

Defining $|\psi(t_A)\rangle\equiv\langle t_A|\Psi\rangle\rangle$ and using Eqs. \eqref{eq:gen} and \eqref{eq:constraint}, we obtain the Schr\"odinger equation
\begin{equation}
    i\hbar \frac{\partial}{\partial t_A} |\psi(t_A)\rangle = H_{\text{eff}}^A |\psi(t_A)\rangle,
    \label{eq:se-a}
\end{equation}
where
\begin{equation}
    H_{\text{eff}}^A \equiv [I + f(X_M)]^{-1} H_M
    \label{eq:h-eff-first}
\end{equation}
is the effective Hamiltonian of system $M$ with respect to clock $A$. Generally speaking, $H_{\text{eff}}^A$ is non-Hermitian since $[I + f(X_M)]^{-1}$ does not always commute with $H_M$. A detailed derivation of the above expression can be seen in the Supplementary Note 1.

It is worth noting that, if a non-interacting clock $B$ external to $M$ is included in the analysis, the only change to the total Hamiltonian of the joint system is the addition of the Hamiltonian of this clock since it does not get coupled to $M$. As a consequence, it follows by direct computation that the effective Hamiltonian from the perspective of clock $B$ is Hermitian. More precisely, it is $H_{\text{eff}}^B = H$, where $H$ is the Hamiltonian in Eq. \eqref{eq:ham-accel-gen}. Moreover, with this remark, we can also verify that our result is in harmony with a quantum field treatment of accelerated clocks in a fixed background spacetime that showed that the time rate of these clocks is affected by their acceleration (and not only by their instant velocity) \cite{lorek2015ideal}. In fact, in our treatment, the time rate of clock $A$ from the perspective of clock $B$ can be calculated as
\begin{equation}
    \frac{d}{dt_B} T_A = \frac{i}{\hbar} [H_{\text{eff}}^B, T_A] = \frac{i}{\hbar} [I + f(X_M)] [H_A, T_A],
\end{equation}
which depends on $f$ and, hence, on the acceleration of clock $A$.

To illustrate and evidence the significance of the results just presented, we discuss the case of two rocket ships $R$ and $S$, each with their own internal clock, as shown in Fig. \ref{fig:rockets}. Initially, we assume that there is no interaction between the rockets nor between each rocket's external degree of freedom with its clock. Thus, if the two rockets are inertial, as represented in Fig. \ref{fig:rockets}(a), the evolution from the perspective of either clock is unitary.

Now, suppose rocket $R$ starts accelerating while rocket $S$ remains inertial, as illustrated in Fig. \ref{fig:rockets}(b). Then, according to the result just presented, the effective dynamics from the perspective of rocket $R$'s clock is, generally, non-unitary. While this is the case, in the derivation of the Hamiltonian in Eq. \eqref{eq:h-eff-first} only the external degrees of freedom of the accelerating clock were taken into consideration. However, it can be readily seen that, if other systems that do not interact with system $M$ and clock $A$ are included in the analysis, their dynamics will be unitary from $A$'s perspective. In fact, their individual Hamiltonians will be added to $H_{\text{eff}}^A$ with a multiplication by the factor $[I + f(X_M)]^{-1}$, which commutes with them. Thus, in the example in Fig. \ref{fig:rockets}(b), the dynamics of rocket $S$ from the perspective of $R$'s clock is governed by a Hermitian Hamiltonian. This is in accordance with a recent analysis that used Fermi-Walker coordinates to study the dynamics according to accelerated clocks of systems that do not interact with them \cite{roura2020gravitational}. There, it was found that their effective evolution is unitary. To conclude the analysis of the scenario in Fig. \ref{fig:rockets}(b), it is noteworthy that the evolution of the systems under consideration from the perspective of rocket $S$'s clock is unitary.

Finally, if both rockets are accelerating, as in Fig. \ref{fig:rockets}(c), the evolution given by either clock is, generally, non-unitary. However, as it will be shown next with an example of gravitationally interacting systems, if a given system interacts with an accelerating one, the effective dynamics from the perspective of the former's clock is, generally, non-Hermitian, even if that system is approximated as inertial.

\textbf{Gravitationally interacting clocks.} We now focus on accelerations due to gravitational interactions between massive systems, each with their own internal clock. This will allow us to reconsider previous results on gravitationally interacting clocks. In these studies, it was shown that gravitational interactions lead to a subtle form of time dilation, although the unitarity of the evolution persists \cite{castro2017entanglement, sonnleitner2018mass, zych2019gravitational, smith2019quantizing, castro2020quantum}. 

In those analyses, it was used that the Newtonian gravitational potential is written as $V(r) = -G m_A m_B/r$ together with the post-Newtonian mass-energy equivalence correction, as we have done to obtain our main result. Then, the gravitational interaction between two clocks $A$ and $B$ was added to the Hamiltonian as a term proportional to the product of their free Hamiltonians, i.e., $\lambda H_A H_B$, where $\lambda = -G/c^4r$.

\begin{figure}
    \centering
    \includegraphics[width=\columnwidth]{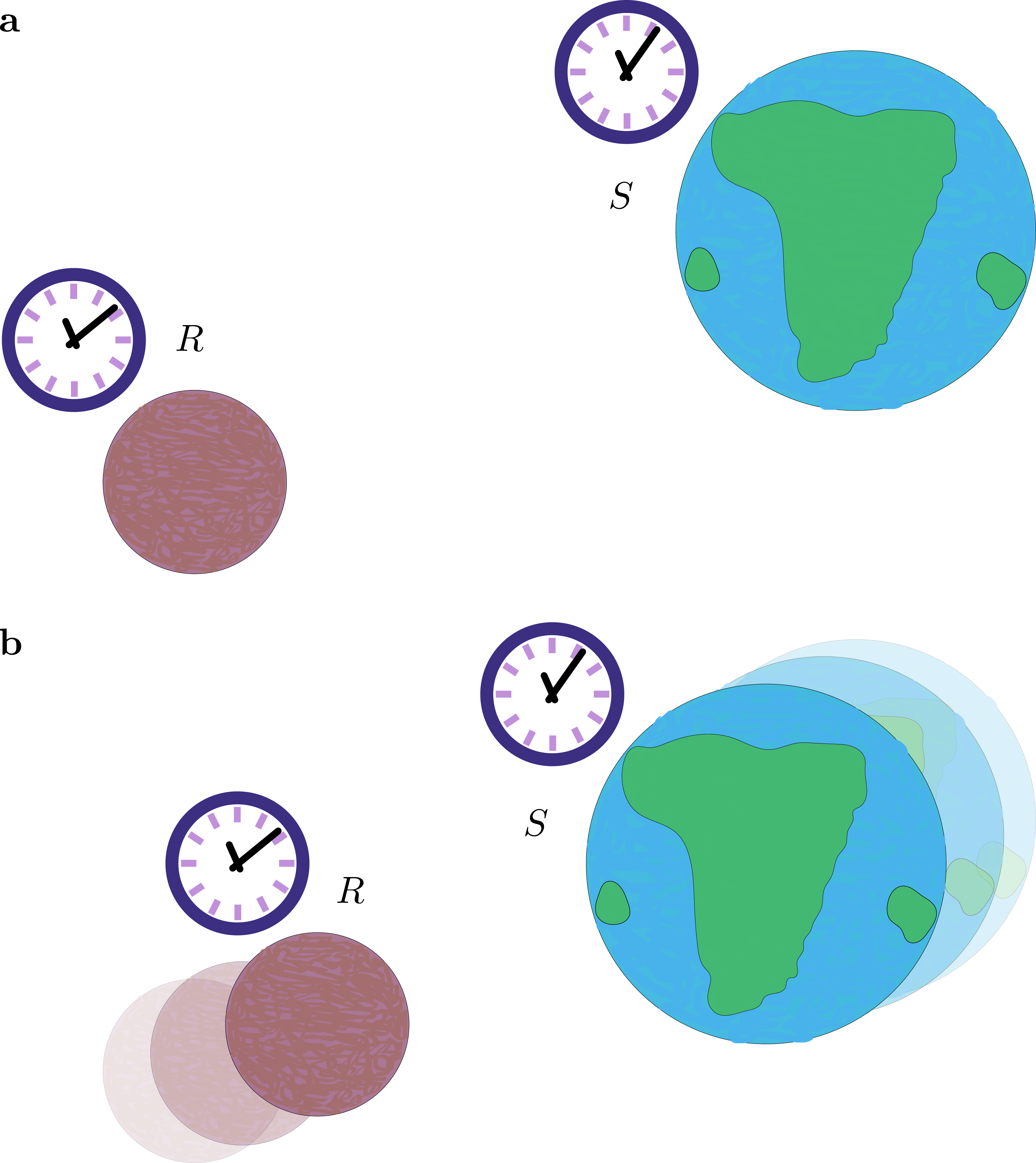}
    \caption{\textbf{Representation of gravitational interaction between two massive systems.} Systems $R$ and $S$ have each their own internal quantum time. As shown in \textbf{a}, if the relative spatial distance between the systems is kept constant, the time evolution from the perspective of either of their clocks is unitary, even though the gravitational interaction causes a type of time dilation in the clocks. However, as illustrated in \textbf{b}, if the relative spatial distance between the systems changes due to the gravitational attraction, the description of their time evolution given by either of their clocks is generally non-unitary. This is the case even if one of the systems is assumed to be much more massive than the other and, therefore, can be approximated as inertial.}
    \label{fig:gravitational-clocks}
\end{figure}

It can be noticed that in these previous works the distance between the clocks was assumed to remain constant. This justifies the fact that, despite the gravitational effects, both clocks were found to yield unitary evolution. In fact, both are inertial frames. Here, however, we allow the relative position of the clocks to change as a result of the gravitational interaction.

More precisely, we assume that clocks $A$ and $B$ are internal degrees of freedom of massive particles $M$ and $N$, respectively. Then, the gravitational potential can be written as $V(x_N-x_M) = -G m_M m_N/|x_N-x_M|$. For simplicity, if $S$ is much more massive than $R$, we can assume that $x_N-x_M \approx x_N - x_0$, where $x_0$ is the initial position of system $M$. Letting the latter vanish, we have $V(x_N) = -G m_M m_N/|x_N|$.

Now, using the mass-energy equivalence, we write $V(X_N) = -G H_A H_B |X_N|^{-1}/2c^4$. For simplicity and to make sure that any change to the ticking of either clock is due to the gravitational interaction between the systems, we assume no other interaction between them. This means that the total Hamiltonian of the composed system is
\begin{equation}
    H = [I + f(X_N, H_B)] H_A + H_B + H_M + H_N,
\end{equation}
where $f(X_N, H_B) = -G H_B |X_N|^{-1}/2c^4$. As a result, the dynamics from the perspective of clock $A$ is given by an expression similar to Eq. \eqref{eq:se-a} with
\begin{equation}
    H_{\text{eff}}^A = [I + f(X_N, H_B)]^{-1} (H_B + H_M + H_N),
    \label{eq:grav-a}
\end{equation}
which is non-Hermitian since $f(X_N, H_B)$ does not commute with $H_N$. This is so in spite of system $S$ being assumed to be approximately inertial. This might seem surprising in view of the example with the rocket ships. However, a crucial aspect in that example is that the rockets had no interaction between them whatsoever. Here, system $S$ interacts with the non-inertial system $R$. Moreover, the effective Hamiltonian from the perspective of clock $B$ is also non-Hermitian, as expected. More precisely,
\begin{equation}
    H_{\text{eff}}^B = [I + f(X_N, H_A)]^{-1} (H_A + H_M + H_N).
    \label{eq:grav-b}
\end{equation}
Details of the derivation of Eqs. \eqref{eq:grav-a} and \eqref{eq:grav-b} can be found in the Supplementary Note 1. Also, the results just discussed are illustrated in Fig. \ref{fig:gravitational-clocks} with $S\equiv A+M$ and $R\equiv B+N$.

It is worth highlighting that gravitationally interacting systems were also considered in a recently introduced spacetime quantum reference frame  \cite{giacomini2021spacetime}. There, the weak-field limit was assumed in order to avoid the problem of ordering of operators. In this limit, the dynamics was found to be unitary. This is also consistent with our results since the non-Hermitian character of the dynamics is accentuated at higher energies. However, since the Hermiticity of the dynamics appears only as an approximation, predictions using this type of approximation might likely deviate from the ones using the non-Hermitian Hamiltonians found here in experiments that are not relatively short.

It is also noteworthy that an analysis of the dynamics of quantum systems in the presence of singularities with different clocks has revealed the manifestation of non-Hermitian dynamics and, more specifically, non-unitary evolution \cite{gielen2022unitarity}. In fact, it was found that, in this scenario, unitarity depends on the choice of the clock --- even if every clock under consideration is a counterpart of good clocks at the classical level. Moreover, it was concluded that the general covariance of general relativity turns out to be incompatible with quantum unitary dynamics.

\textbf{Parametrization by time states.} A limitation of the approach used in ``Accelerating clock frames'' is the requirement that the acceleration is parametrized by the position of the center-of-mass of the system. In the case of a single spacial dimension, this implies that the motion is unidirectional. On the one hand, this approach is useful to establish connections with fundamental interactions, like we have done with gravitational interactions, since these interactions are typically given as a function of the position. On the other hand, it is possible to avoid these limitations by parametrizing the acceleration with time states.

Using this approach, with the correction due to the acceleration, the Hamiltonian of the system becomes
\begin{equation}
    H = H_A + H_M + m \int dt_A g(t_A) |t_A\rangle \langle t_A|.
\end{equation}
If we use the mass-energy equivalence in a similar manner to the above, we are faced with an ordering issue since $H_A$ does not commute with $\int dt_A g(t_A) |t_A\rangle \langle t_A|$ in general.

If a non-symmetric order is chosen, then, the resulting dynamics should generally be non-unitary from the perspective of external clocks that do not interact with clock $A$. Because of this, we choose the Weyl ordering and obtain the Hamiltonian in Eq. \eqref{eq:hamiltonian-generic} with
\begin{equation}
    H_{\text{int}} = \frac{1}{2} \int dt_A g(t_A) \left[H_A |t_A\rangle \langle t_A| + |t_A\rangle \langle t_A| H_A \right].
\end{equation}
As already discussed, this gives rise to the dynamics in Eq. \eqref{eq:schrod-general}. To see that this dynamics is generally non-unitary, we show in the Supplementary Note 2 that, in the case of an ideal clock, it reduces to Eq. \eqref{eq:se-a} with
\begin{equation}
    H_{\text{eff}}^A = [1-g(t_A)]^{-1} \left[H_M - \frac{i\hbar}{2} g'(t_A) I\right]
\end{equation}
and, moreover, if $H_M = P_M^2/2m$ and $|\psi(0)\rangle = (2\Delta^2/\pi)^{1/4} \int dp_M \; e^{-\Delta^2 p_M^2} |p_M\rangle$, we have
\begin{equation}
    \langle\psi(t_A)|\psi(t_A)\rangle = e^{-\int_0^{t_A} dt'_A \; [1-g(t'_A)]^{-1} g'(t_A)},
\end{equation}
which, generally, is not constant in time.

\section*{Discussion}

We have studied how accelerations of massive particles lead to the emergence of non-Hermitian dynamics and even non-unitarity from the perspective of quantum clocks internal to them. These results come as a consequence of the coupling between external and internal degrees of freedom of a system, which include a clock system (associated with the system's proper time). This is a general feature arising from the Page and Wootters formalism, not relying on specific implementations of the clocks or even on them being ideal. It contrasts, for instance, with the already discussed result in Ref. \cite{gielen2022unitarity} and also with an analysis of quantum clocks in superpositions of different states of motion in relativistic scenarios \cite{khandelwal2020universal}. In the latter, it was found that even the average behavior of the clock can be affected by its preparation. Nevertheless, our result reveals a generic feature that should be present regardless of any specific characteristic of the clock at hand, as seen in our derivations.

By the equivalence principle, accelerating massive particles are equivalent to systems under gravitational forces. To evidence this, we have conducted an explicit analysis of gravitationally interacting systems. This allowed us to explain why non-Hermitian dynamics was not observed in previous theoretical treatments of the problem \cite{castro2017entanglement, sonnleitner2018mass, zych2019gravitational, smith2019quantizing, castro2020quantum}. Moreover, the relation between our results and gravitational effects is particularly emblematic: since every system interacts through gravity, our results suggest that there is no clock frame in the Page and Wootters framework for which the effective dynamics is exactly Hermitian. In other words, since, ultimately, any system can be addressed as being inertial only up to a certain order, the results presented here suggest that unitarity can only be recovered as an approximation in an eventual quantum theory incorporating gravitation. Hence, the time evolution of a system should be, in general, non-unitary in those theories.

The results presented here can be assimilated in two different ways. In one way, it is possible to conclude that a non-unitary evolution will indeed be a fundamental characteristic of relativistic quantum theories with an operational approach to time --- and, in particular, to a yet-to-be-constructed consistent theory of quantum gravity. In this case, it is necessary to develop an understanding of the physical meaning of such evolution. For instance, by the construction of the state $|\psi(t_A)\rangle$ according to Page and Wootters' recipe in their framework, the state $|\psi(t_A)\rangle$ in Eq. \eqref{eq:se-a} is a vector (i.e., a pure state) for every $t_A$. The difference between unitary and non-unitary dynamics in this context is, then, that the norm of the vector changes in time within the latter. Knowing that, how can the usual operational meaning of quantum mechanics be recovered? More, since the Schr\"odinger and the Heisenberg pictures are unitarily equivalent, how can the Heisenberg picture be recovered in this case?

To recover the probabilistic notion, one can, in principle, divide the results obtained with standard calculations by the instantaneous norm of the state vector. In fact, the use of normalized vectors in standard quantum mechanics is a convenience in view of the probabilistic operational meaning of the state vector combined with the fact that the norm does not change during a unitary dynamics. Nevertheless, in general, we can, for instance, calculate the expected value of an operator $O$ for a system in the (possibly non-normalized) state $|\psi\rangle$ as $\langle O \rangle = \langle\psi|O|\psi\rangle/\langle\psi|\psi\rangle$.

However, this does not allow the ``reconstruction'' of the Heisenberg picture. A possible solution to the issues raised here that includes the latter may lie within a method to treat non-Hermitian Hamiltonians introduced by Dirac \cite{dirac1942bakerian} and further studied by Pauli \cite{pauli1943dirac} and others \cite{lee1969negative, scholtz1992quasi, ju2019non}. The method consists of introducing a new metric to the Hilbert space of the system, which modifies its inner product. This new metric should be such that the new norm of the vector is kept constant throughout its evolution. This, however, comes at a price: the choice of a new metric is not unique and, most disturbingly, it is not guaranteed \textit{a priori} that there always exist a positive-definite metric. This means that the theory may have states with negative norms, known as ``ghost states'' since they do not have an operational meaning. For instance, in the case of $\mathcal{PT}$-symmetric systems, the metric induced by the $\mathcal{PT}$-symmetry is indefinite. However, if the $\mathcal{PT}$-symmetry is unbroken, these systems have a third symmetry that can be used to construct a positive-definite metric \cite{bender2002complex}.

Then, one may question what are the physical consequences of the change of inner product. While there is much to be investigated in this regard, some hints may be found in the literature of (non-Hermitian) $\mathcal{PT}$-symmetric quantum mechanics \cite{bender1998real, lee1954some, wu1959ground, brower1978critical, fisher1978yang, bender2002complex}. For instance, it is known that $\mathcal{PT}$-symmetric systems can evolve faster than Hermitian ones \cite{bender2007faster, zheng2013observation}. Therefore, quantum bounds that rely on inner products may, in general, be modified.

The other way to look at the results presented in this article consists of seeing them as a limitation of the Page and Wootters framework. Since there will be no perspective from which the evolution is Hermitian in a scenario where every system interacts through gravity, the issue of non-Hermitian dynamics does not appear to be necessarily related to the manner the change of perspective is implemented. Instead, it seems that a reevaluation of the foundations of the framework will be required when modifying/extending it to restore Hermiticity.

In either case, the present work reveals challenges for devising operational approaches to time in relativistic quantum theories. These challenges, in turn, bring new research directions that may lead to a better understanding of time and relativistic structures in quantum mechanics.

\section*{Acknowledgements}

We thank the anonymous referees for their constructive comments that helped to highly improve this work. We also thank Magdalena Zych for insightful comments on a previous version of this work. This research was supported by the Fetzer-Franklin Fund of the John E. Fetzer Memorial Trust and by Grant No. FQXi-RFP-CPW-2006 from the Foundational Questions Institute and Fetzer Franklin Fund, a donor-advised fund of Silicon Valley Community Foundation. E.C. was supported by the Israeli Innovation Authority under Projects No. 70002 and No. 73795, by the Pazy Foundation, by the Israeli Ministry of Science and Technology, and by the Quantum Science and Technology Program of the Israeli Council of Higher Education. Y.A. thanks the Federico and Elvia Faggin Foundation for support.

\section*{Data Availability}

Data sharing not applicable to this article as no datasets were generated or analysed.

\section*{Author contributions}

I.L.P., A.T., B.Y.P., E.C., and Y.A. contributed to the preparation of this article.

\section*{Competing interests}

The authors declare no competing interests.

\bibliography{citations}

\onecolumngrid

\section*{Supplementary Note 1: Derivation of effective Hamiltonians}

Here, we provide details for the derivation of the Hamiltonians in Eqs. (11), (14), and (15).

First, we derive $H_{\text{eff}}^A$ in Eq. (11). Defining $|\psi(t_A)\rangle\equiv\langle t_A|\Psi\rangle\rangle$ and using Eqs. (2), (3), and (9), we obtain
\begin{equation}
    \begin{aligned}
        0 &= \langle t_A|H|\Psi\rangle\rangle \\
          &= \langle t_A |\left\{\left[I+f(X_M)\right]H_A + H_M \right\}|\Psi\rangle\rangle \\
          &= \left[I+f(X_M)\right]\left(\langle t_A |H_A\right) |\Psi\rangle\rangle + H_M \langle t_A|\Psi\rangle\rangle \\
          &= \left[I+f(X_M)\right]\left(-i\hbar\frac{\partial}{\partial t_A}\langle t_A |\right) |\Psi\rangle\rangle + H_M |\psi(t_A)\rangle \\
          &= -i\hbar \left[I+f(X_M)\right] \frac{\partial}{\partial t_A}|\psi(t_A)\rangle + H_M |\psi(t_A)\rangle.
    \end{aligned}
\end{equation}
Assuming that $I+f(X_M)$ is invertible and reorganizing the terms, we obtain Eq. (10) with $H_{\text{eff}}^A$ given by Eq. (11). In the case we are considering, we assume the particle is moving in a single direction since its acceleration can be parameterized by its position. Then, it should be possible to use systems of coordinates such that $I+f(X_M)$ is invertible everywhere. Yet, it should be noted that, whenever this expression is not invertible, an equation similar to Eq. (10) continues to hold. However, in this expression, the term $I+f(X_M)$ should remain multiplying the time derivative.

Now, a similar derivation leads to the Hamiltonian in Eq. (14). Defining $|\psi(t_A)\rangle\equiv\langle t_A|\Psi\rangle\rangle$ and using Eqs. (2), (3), and (13), we obtain
\begin{equation}
    \begin{aligned}
        0 &= \langle t_A |\left\{[I + f(X_N, H_B)] H_A + H_B + H_M + H_N \right\}|\Psi\rangle\rangle \\
          &= \left[I + f(X_N, H_B)\right]\left(\langle t_A |H_A\right) |\Psi\rangle\rangle + H_B \langle t_A|\Psi\rangle\rangle + H_M \langle t_A|\Psi\rangle\rangle + H_N \langle t_A|\Psi\rangle\rangle \\
          &= \left[I + f(X_N, H_B)\right]\left(-i\hbar\frac{\partial}{\partial t_A}\langle t_A |\right) |\Psi\rangle\rangle + \left(H_B + H_M + H_N\right) |\psi(t_A)\rangle \\
          &= -i\hbar \left[I + f(X_N, H_B)\right] \frac{\partial}{\partial t_A}|\psi(t_A)\rangle + \left(H_B + H_M + H_N\right) |\psi(t_A)\rangle.
    \end{aligned}
\end{equation}
Assuming that $I + f(X_N, H_B)$ is invertible and reorganizing the terms, we obtain Eq. (10) with $H_{\text{eff}}^A$ given by Eq. (14).

Finally, the same reasoning leads to Eq. (15). It is just necessary to observe that the Hamiltonian $H$ can be also written as
\begin{equation}
    H = H_A + [I + f(X_N, H_A)] H_B + H_M + H_N.
\end{equation}

\section*{Supplementary Note 2: Example of non-unitary dynamics with ideal clock}

Here, we show that the Hamiltonian in Eq. (5) with $H_{\text{int}}$ given by Eq. (17) is generally non-unitary. For simplicity, we assume the clock is ideal. In this case, we first observe that
\begin{equation}
    \int dt_A g(t_A) |t_A\rangle \langle t_A| = g(T_A).
\end{equation}
Then, we can write
\begin{equation}
    H_{\text{int}} = \frac{1}{2} \left[H_A g(T_A) + g(T_A) H_A\right].
\end{equation}

Since the pair $T_A$ and $H_A$ is analogous to position and momentum for ideal clocks, we have $[H_A, g(T_A)] = -i\hbar g'(T_A)$, where $g$ was assumed to be an analytic function and $g'$, its first derivative. As a result,
\begin{equation}
    H_{\text{int}} = g(T_A) H_A - \frac{i\hbar}{2} g'(T_A).
\end{equation}
Then, recalling that $T_A|t_A\rangle = t_A|t_A\rangle$ for ideal clocks, Eq. (7) in the case of interest becomes
\begin{equation}
    \begin{aligned}
        i\hbar \frac{\partial}{\partial t_A} |\psi(t_A)\rangle &= H_M |\psi(t_A)\rangle + \int dt'_A \ K(t_A,t'_A)|\psi(t'_A)\rangle \\
           &= H_M |\psi(t_A)\rangle + i\hbar \left[g(t_A) \frac{\partial}{\partial t_A} - \frac{1}{2} g'(t_A)\right] |\psi(t_A)\rangle.
    \end{aligned}
\end{equation}
Reorganizing the terms and assuming that $1-g(t_A)$ is invertible we obtain the Hamiltonian in Eq. (10) with $H_{\text{eff}}^A$ given by Eq. (18).

If $H_M = P_M^2/2m$ and the initial state of system $M$ is $|\psi(0)\rangle = (2\Delta^2/\pi)^{1/4} \int dp_M \; e^{-\Delta^2 p_M^2} |p_M\rangle$, we have
\begin{equation}
    |\psi(t_A)\rangle = \left(\frac{2\Delta^2}{\pi}\right)^{1/4} e^{-\int_0^{t_A} dt'_A \; [1-g(t'_A)]^{-1} g'(t_A)/2} \int dp_M \; e^{-i \int_0^{t_A} dt'_A \; [1-g(t'_A)]^{-1} p_M^2/2m\hbar} e^{-\Delta^2 p_M^2} |p_M\rangle
\end{equation}
and, moreover, the result in Eq. (19).

\end{document}